\documentstyle[12pt]{article}
\begin{document}

\makeatletter
\@addtoreset{equation}{section}
\def\theequation{\thesection.\arabic{equation}}
\makeatother

\begin{flushright}{UT-885}
\end{flushright}
\vskip 0.5 truecm

\begin{center}
{\large{\bf Quantum and Classical Gauge Symmetries in a Modified
Quantization Scheme }}
\end{center}
\vskip .5 truecm
\centerline{\bf Kazuo Fujikawa and Hiroaki Terashima}
\vskip .4 truecm
\centerline {\it Department of Physics,University of Tokyo}
\centerline {\it Bunkyo-ku,Tokyo 113,Japan}
\vskip 0.5 truecm

\begin{abstract}
The use of the mass term as a gauge fixing term has been 
studied by Zwanziger, Parrinello 
and Jona-Lasinio, which is related to the non-linear gauge 
$A_{\mu}^{2}=\lambda$ of Dirac and Nambu in the large mass limit.
We have recently shown that this modified quantization scheme  is 
in fact identical to the 
conventional {\em local} Faddeev-Popov formula {\em without} 
taking the
 large mass limit, if one takes into account the variation of 
the gauge field along the entire gauge orbit and if the Gribov
complications can be ignored. This 
suggests that the classical massive vector theory, for example, 
is interpreted in a more flexible manner either as a gauge 
invariant theory with a gauge fixing term added, or as a 
conventional massive non-gauge theory.  
As for massive gauge particles, the Higgs 
mechanics, where the  mass term is gauge invariant, has a
more intrinsic meaning.
 It is suggested to extend the notion of quantum 
gauge symmetry (BRST symmetry) not only to classical gauge theory 
but also to a wider class of  theories whose gauge symmetry is 
broken by some extra terms in the classical action. We comment on 
the implications of this extended notion of quantum gauge 
symmetry.
\end{abstract}

\section{Introduction}

We have recently shown\cite{fujikawa} that the modified
quantization scheme\cite{zwanziger}\cite{jona-lasinio}
\begin{equation}
\int{\cal D}A_{\mu}\{\exp[-S_{YM}(A_{\mu})-
\int f(A_{\mu})dx]/\int{\cal 
D}g\exp[-\int f(A_{\mu}^{g})dx]\}
\end{equation}
with a gauge non-invariant term $f(A_{\mu})$, for example,
\begin{equation}  
f(A_{\mu})=(m^{2}/2)(A_{\mu})^{2}
\end{equation}
is identical at least  to the 
conventional local Faddeev-Popov formula\cite{faddeev}
\begin{eqnarray}
&&\int{\cal D}A_{\mu}\{\delta(D^{\mu}\frac{\delta f(A_{\nu})}
{\delta A_{\mu}})/\int {\cal D}g\delta(D^{\mu}
\frac{\delta f(A_{\nu}^{g})}
{\delta A_{\mu}^{g}})\}\exp[-S_{YM}(A_{\mu})]\nonumber\\
&&=\int{\cal D}A_{\mu}\delta(D^{\mu}\frac{\delta f(A_{\nu})}
{\delta A_{\mu}})\det\{\delta[D^{\mu}
\frac{\delta f(A_{\nu}^{g})}{\delta A_{\mu}^{g}}]/\delta g\}
\exp[-S_{YM}(A_{\mu})]
\end{eqnarray}
{\em without} taking the large mass limit, if one takes into 
account the variation of the gauge field along 
the entire gauge orbit parametrized by the gauge parameter $g$. 
Here the operator $D^{\mu}\frac{\delta f(A_{\nu})}
{\delta A_{\mu}}(x)$ is defined by an infinitesimal gauge 
transformation as 
\begin{equation}
\int dx f(A_{\nu}+D_{\nu}\omega)=\int dx f(A_{\nu})
-\int dx \omega(x)D^{\mu}\frac{\delta f(A_{\nu})}{\delta A_{\mu}}
(x).
\end{equation}
The above  equivalence was discussed 
in\cite{fujikawa} in connection with the analysis of the so-called
Gribov problem\cite{gribov}, and the above formula is valid if 
the Gribov-type complications are ignored such as in perturbative
approach. We, however,  note that the choice
of the gauge fixing function $f(A_{\mu})$ is rather general but 
not completely arbitrary since certain technical conditions need 
to be satisfied in the above proof of the 
equivalence\cite{fujikawa}.

One can confirm that the starting expression (1.1) is formally
identical to the extraction of the gauge volume from the naive
path integral measure,
\begin{equation}
\int\frac{{\cal D}A_{\mu}}{vol(g)}\exp[-S_{YM}(A_{\mu})]
\end{equation}
and thus the above equivalence is not 
unexpected\cite{zwanziger}\cite{jona-lasinio}. A remarkable
aspect is that the apparently non-local expression (1.1) in fact 
defines a local and thus unitary theory (1.3)\cite{fujikawa}. 
 
We emphasize that the above equivalence is valid for any value of 
the mass parameter $m^{2}$, for example. The large mass limit, 
where the equivalence to the conventional local formula was 
analyzed in the past\cite{zwanziger}\cite{jona-lasinio}, is 
formally related to the non-linear gauge 
\begin{equation}
A_{\mu}^{2}=\lambda=const.,
\end{equation}
of Dirac\cite{dirac} and Nambu\cite{nambu} in the limit
$\lambda=0$. Nambu used the above gauge to analyze the 
possible spontaneous breakdown of Lorentz symmetry. In his
treatment, the limit $\lambda=0$ is singular, and thus the 
present formulation is not quite convenient for an analysis of 
the possible breakdown of Lorentz symmetry. See, for example, 
\cite{fujikawa0} for the past analyses of 
the above non-linear gauge. 

In this paper, we comment on the possible implications of the 
above equivalence between (1.1) and the local BRST invariant 
theory (1.3) in a general context of 
quantum gauge symmetry, namely, BRST symmetry\cite{brs}, which 
controls the analyses of renormalization and unitarity. The 
modified quantization scheme is flexible in choosing  
gauge fixing functions $f(A_{\mu})$, and we argue that from a 
view point of quantum gauge symmetry there 
is no intrinsic difference between the classical theory with some 
extra terms such as a mass term, which break gauge symmetry, and
the classical gauge theory whose gauge symmetry is broken by a 
gauge fixing term. In particular, the classical 
massive Yang-Mills theory is more flexibly interpreted either as
a gauge fixed version of  pure Yang-Mills theory in the modified 
quantization scheme, or as the conventional massive non-gauge 
theory.

Our basic argument is a possible {\em re-interpretation} of 
various classical Lagrangians, but it could also be understood as
 a broader choice of path integral measure than in the 
conventional analysis starting with the Dirac bracket.
The spirit of our approach is probably illustrated by considering
 the two-dimensional field theory coupled to gravity
\begin{equation}
{\cal L}=-\frac{1}{2}\sqrt{g}g^{\mu\nu}\partial_{\mu}X^{a}(x)
\partial_{\nu}X^{a}(x).
\end{equation}
In the conformal gauge defined by 
\begin{equation}
g_{\mu\nu}=\delta_{\mu\nu}\rho(x)
\end{equation}
the metric $\rho(x)$ decouples from the above action, and one may
define the path integral\cite{sakita}
\begin{equation}
\int{\cal D}X^{a}\exp[-\int \frac{1}{2}\partial_{\mu}X^{a}(x)
\partial^{\mu}X^{a}(x)dx].
\end{equation}
On the other hand, it is well known that a  
reparametrization invariant path integral measure leads to
\cite{polyakov}\cite{fujikawa1} 
\begin{equation}
\int{\cal D}X^{a}{\cal D}\sqrt{\rho}\exp\{-\int \frac{1}{2}
\partial_{\mu}X^{a}(x)\partial^{\mu}X^{a}(x)dx-\frac{26-d}{24\pi}
\int[\partial_{\mu}\ln\rho\partial^{\mu}\ln\rho+\mu^{2}\rho]dx\}
\end{equation}  
where $d$ stands for the number of variables $X^{a}$.
The last term in the action, which is regarded as a Wess-Zumino
term, is known as the Liouville action.
In both of these expressions, we neglected the Faddeev-Popov 
ghosts, which should be included in a consistent quantization
\cite{fujikawa1}.
The above example, which appears in the first quantization of 
string theory, illustrates that the choice of path integral 
measure has more freedom than in the naive prescription 
starting with Dirac brackets. It is also known that many 
interesting physical examples can be quantized consistently 
only when one adds a suitable Wess-Zumino term, which may be 
regarded as a modified choice of measure. See, for example,
the quantization of anomalous gauge theory in 
2-dimensions\cite{jackiw} and 
the quantization of supermembrane in\cite{bergshoeff}.

\section{Abelian example}

We first briefly illustrate the proof\cite{fujikawa} of  the 
above equivalence of (1.1) and (1.3) by using an 
example of  Abelian gauge theory, 
\begin{equation}
S_{0}=-\frac{1}{4}\int dx
(\partial_{\mu}A_{\nu}-\partial_{\nu}A_{\mu})^{2}
\end{equation}
for which we can work out 
everything explicitly. In this note we exclusively work on  
Euclidean theory with metric convention $g_{\mu\nu}=(1,1,1,1)$. 
Note that there is no Gribov complications
in the Abelian theory at least in a continuum formulation. 
 As a simple and useful example, we choose
the gauge fixing function\cite{zwanziger}\cite{jona-lasinio}
\begin{equation}
f(A)\equiv \frac{1}{2}A_{\mu}A_{\mu}
\end{equation}
and 
\begin{equation}
D_{\mu}(\frac{\delta f}{\delta A_{\mu}})=\partial_{\mu}A_{\mu}.
\end{equation}
Our claim above  suggests the relation
\begin{eqnarray}
Z&=&\int {\cal D}A^{\omega}_{\mu}\{e^{-S_{0}(A^{\omega}_{\mu})-
\int dx \frac{1}{2}(A^{\omega}_{\mu})^{2}}/\int {\cal D}h 
e^{-\int dx 
\frac{1}{2}(A^{h\omega}_{\mu})^{2}}\}\nonumber\\
&=&\int {\cal D}A^{\omega}_{\mu}{\cal D}B{\cal D}\bar{c}{\cal D}c 
e^{-S_{0}(A^{\omega}_{\mu})+ \int[-iB\partial_{\mu}
A^{\omega}_{\mu}+ 
\bar{c}(-\partial_{\mu}\partial_{\mu})c]dx }
\end{eqnarray}
where the variable $A^{\omega}_{\mu}$ stands for the field
variable obtained from $A_{\mu}$ by a gauge transformation
parametrized by the gauge orbit parameter $\omega$. 
To establish this result, we first evaluate 
\begin{eqnarray}
&&\int {\cal D}h e^{-\int dx \frac{1}{2}(A^{h\omega}_{\mu})^{2}} 
\nonumber\\
&&=\int {\cal D}h e^{-\int dx \frac{1}{2}(A^{\omega}_{\mu}
+\partial_{\mu}
h)^{2}} \nonumber\\
&&=\int {\cal D}h e^{-\int dx 
\frac{1}{2}[(A^{\omega}_{\mu})^{2}-2(\partial_{\mu}
A^{\omega}_{\mu})h + 
h(-\partial_{\mu}\partial_{\mu})h]} 
\nonumber\\
&&=\int {\cal D}B\frac{1}{det\sqrt{-\partial_{\mu}
\partial_{\mu}}} 
e^{-\int dx \frac{1}{2}[(A^{\omega}_{\mu})^{2}
-2(\partial_{\mu}A^{\omega}
_{\mu})\frac{1}{\sqrt{-\partial_{\mu}\partial_{\mu}}}B + 
B^{2}]}\nonumber\\
&&=\frac{1}{det \sqrt{-\partial_{\mu}\partial_{\mu}}}
e^{-\int dx \frac{1}
{2}(A^{\omega}_{\mu})^{2}+\frac{1}{2}\int\partial_{\mu}
A^{\omega}_{\mu}\frac{1}{-\partial_{\mu}\partial_{\mu}}
\partial_{\nu}A^{\omega}_{\nu}dx} 
\end{eqnarray}
where we defined $\sqrt{-\partial_{\mu}\partial_{\mu}}h=B$.
 Thus
\begin{eqnarray}
Z&=&\int{\cal D}A^{\omega}_{\mu}\{det 
\sqrt{-\partial_{\mu}\partial_{\mu}}\}e^{-S_{0}(A^{\omega}_{\mu})
-\frac{1
}{2}\int\partial_{\mu}A^{\omega}_{\mu}\frac{1}{-\partial_{\mu}
\partial_{\mu}}
\partial_{\nu}A^{\omega}_{\nu}dx}
\nonumber\\
&=&\int{\cal D}A^{\omega}_{\mu}{\cal D}B{\cal D}\bar{c}{\cal D}c
e^{-S_{0}(A^{\omega}_{\mu})-\frac{1}{2}\int B^{2}dx +\int
[-iB\frac{1}{\sqrt{-\partial_{\mu}\partial_{\mu}}}
\partial_{\mu}A^{\omega
}_{\mu}+\bar{c}\sqrt{-\partial_{\mu}\partial_{\mu}}c]dx}
\end{eqnarray}
which is invariant under the BRST transformation
\begin{eqnarray}
&&\delta A_{\mu}^{\omega}=i\lambda\partial_{\mu}c\nonumber\\
&&\delta c=0\nonumber\\
&&\delta\bar{c}=\lambda B\nonumber\\
&&\delta B=0
\end{eqnarray}
with a Grassmann parameter $\lambda$. Note the appearance of the
imaginary factor $i$ in the term
$iB\frac{1}{\sqrt{-\partial_{\mu}\partial_{\mu}}}
\partial_{\mu}A^{\omega}
_{\mu}$ 
in (2.6). The expression (2.6) has been derived in 
\cite{zwanziger}\cite{jona-lasinio}. To show that we in fact 
obtain a local theory, we need to go one more step further.

We next rewrite the expression (2.6) as
\begin{eqnarray}
&&\int{\cal D}A^{\omega}_{\mu}{\cal D}B{\cal D}\Lambda{\cal 
D}\bar{c}{\cal D}c 
\delta(\frac{1}{\sqrt{-\partial_{\mu}\partial_{\mu}}}
\partial_{\mu}A^{\omega}_{\mu}-\Lambda)e^{-S_{0}(A^{\omega}_{\mu})
-\frac{1}{2}\int(B^{2}+2i\Lambda B)dx 
+\int\bar{c}\sqrt{-\partial_{\mu}\partial_{\mu}}cdx}\nonumber\\
&&=\int{\cal D}A^{\omega}_{\mu}{\cal D}\Lambda{\cal D}\bar{c}
{\cal D}c 
\delta(\frac{1}{\sqrt{-\partial_{\mu}\partial_{\mu}}}
\partial_{\mu}
A^{\omega}_{\mu}-\Lambda)e^{-S_{0}(A^{\omega}_{\mu})-\frac{1}{2}
\int\Lambda^{2}dx +\int\bar{c}\sqrt{-\partial_{\mu}
\partial_{\mu}}cdx}.
\end{eqnarray}
We note that we can compensate any variation of 
$\delta\Lambda$ by a suitable change of gauge parameter 
$\delta\omega$ inside the $\delta$-function as 
\begin{equation}
\frac{1}{\sqrt{-\partial_{\mu}\partial_{\mu}}}\partial_{\mu}
\partial_{\mu
}\delta\omega=\delta\Lambda.
\end{equation}
By a repeated application of infinitesimal gauge transformations 
combined with the invariance of the path integral measure under 
these gauge transformations, we can re-write the formula (2.8) 
as 
\begin{eqnarray}
&&\int{\cal D}A^{\omega}_{\mu}{\cal D}\Lambda{\cal D}\bar{c}
{\cal D}c 
\delta(\frac{1}{\sqrt{-\partial_{\mu}\partial_{\mu}}}
\partial_{\mu}
A^{\omega}_{\mu})e^{-S_{0}(A^{\omega}_{\mu})-\frac{1}{2}
\int\Lambda^{2}dx 
+\int\bar{c}\sqrt{-\partial_{\mu}\partial_{\mu}}cdx}\nonumber\\
&&=\int{\cal D}A^{\omega}_{\mu}{\cal D}\bar{c}{\cal D}c 
\delta(\frac{1}{\sqrt{-\partial_{\mu}\partial_{\mu}}}
\partial_{\mu}
A^{\omega}_{\mu})e^{-S_{0}(A^{\omega}_{\mu})+\int\bar{c}
\sqrt{-\partial_{\mu}\partial_{\mu}}cdx}\nonumber\\
&&=\int{\cal D}A^{\omega}_{\mu}{\cal D}B{\cal D}\bar{c}{\cal D}c 
e^{-S_{0}(A^{\omega}_{\mu})+\int[-iB\frac{1}{\sqrt{-\partial_{\mu}
\partial_{\mu}}}\partial_{\mu}A^{\omega}_{\mu}+\bar{c}
\sqrt{-\partial_{\mu}\partial_{\mu}}c]dx}\nonumber\\
&&=\int{\cal D}A^{\omega}_{\mu}{\cal D}B{\cal D}\bar{c}{\cal D}c 
e^{-S_{0}(A^{\omega}_{\mu})+\int[-iB\partial_{\mu}
A^{\omega}_{\mu}+\bar{c
}(-\partial_{\mu}\partial_{\mu})c]dx}.
\end{eqnarray}
In the last stage of this equation, we re-defined the 
{\em auxiliary} variables $B$ and $\bar{c}$ as 
\begin{eqnarray}
&&B\rightarrow B\sqrt{-\partial_{\mu}\partial_{\mu}}\nonumber\\
&&\bar{c}\rightarrow \bar{c}\sqrt{-\partial_{\mu}\partial_{\mu}} 
\end{eqnarray}
which is consistent with BRST symmetry and leaves the path 
integral measure invariant. We have thus established the desired 
result (2.4). We emphasize that the integral over the entire 
gauge orbit, as is illustrated in (2.9), is crucial to obtain a 
local expression (2.10)\cite{fujikawa}.

It is shown that this procedure works for the non-Abelian case 
also\cite{fujikawa}, though the actual procedure is much more 
involved and implicit,  if the (ill-understood) Gribov-type 
complications can be ignored such as in perturbative calculations.

\section{No massive gauge fields?}

In the classical level, we traditionally consider
\begin{equation}
{\cal L}= -\frac{1}{4}
(\partial_{\mu}A_{\nu}-\partial_{\nu}A_{\mu})^{2}-\frac{1}{2}
m^{2}A_{\mu}A^{\mu}
\end{equation}
as a Lagrangian for a massive vector theory, and 
\begin{equation}
{\cal L}= -\frac{1}{4}
(\partial_{\mu}A_{\nu}-\partial_{\nu}A_{\mu})^{2}-\frac{1}{2}
(\partial_{\mu}A^{\mu})^{2}
\end{equation}
as an effective Lagrangian for Maxwell theory with a Feynman-type 
gauge fixing term added. The physical meanings of these two 
Lagrangians are thus completely different. 

However, the analysis in Section 2 shows that the Lagrangian 
(3.1) could in fact be regarded as a gauge fixed Lagrangian of 
{\em massless} Maxwell field in quantized theory.  To be 
explicit, by using (2.4), the Lagrangian (3.1) may be regarded as
 an effective Lagrangian in 
\begin{eqnarray}
Z&=&\int {\cal D}A^{\omega}_{\mu}\{e^{\int dx[ -\frac{1}{4}
(\partial_{\mu}A_{\nu}-\partial_{\nu}A_{\mu})^{2}-\frac{1}{2}
m^{2}A_{\mu}^{\omega}A^{\omega\mu}]}
/\int {\cal D}h e^{-\int dx
\frac{m^{2}}{2}(A^{h\omega}_{\mu})^{2}}\}\nonumber\\
&=&\int{\cal D}A^{\omega}_{\mu}{\cal D}B{\cal D}\bar{c}{\cal D}c 
e^{\int dx [-\frac{1}{4}(\partial_{\mu}A_{\nu}-
\partial_{\nu}A_{\mu})^{2}-iB\partial_{\mu}A^{\omega}_{\mu}+ 
\bar{c}(-\partial_{\mu}\partial_{\mu})c]dx}
\end{eqnarray}
where we absorbed the factor $m^{2}$ into the definition of 
$B$ and $\bar{c}$.

One can also analyze (3.2) by defining 
\begin{equation}
f(A_{\mu})\equiv\frac{1}{2}(\partial_{\mu}A^{\mu})^{2}
\end{equation}
in the modified quantization scheme (1.1). The equality of 
(1.1) and (1.3) then gives
\begin{eqnarray}
&&\int{\cal D}A_{\mu}\delta(D^{\mu}\frac{\delta f(A_{\nu})}
{\delta A_{\mu}})\det\{\delta[D^{\mu}
\frac{\delta f(A_{\nu}^{g})}{\delta A_{\mu}^{g}}]/\delta g\}
\exp[-S_{0}(A_{\mu})]\nonumber\\
&&=\int{\cal D}A_{\mu}\delta(\partial_{\nu}\partial^{\nu}
(\partial^{\mu}A_{\mu}))\det[\partial_{\nu}\partial^{\nu}
\partial_{\mu}\partial^{\mu}]
\exp[-S_{0}(A_{\mu})]\\
&&=\int{\cal D}A_{\mu}{\cal D}B{\cal D}\bar{c}{\cal D}c
\exp\{-S_{0}(A_{\mu})
+\int dx[-iB\partial_{\nu}\partial^{\nu}(\partial^{\mu}A_{\mu})
-\bar{c}(\partial_{\nu}\partial^{\nu}
\partial_{\mu}\partial^{\mu})c]\}\nonumber
\end{eqnarray}
After the re-definition of {\em auxiliary}
variables,
\begin{equation}
B\partial_{\nu}\partial^{\nu} \rightarrow B,\ \ \ \
\bar{c}\partial_{\nu}\partial^{\nu} \rightarrow \bar{c}
\end{equation}
which preserves BRST symmetry, (3.5) becomes
\begin{equation}
\int{\cal D}A_{\mu}{\cal D}B{\cal D}\bar{c}{\cal D}c
\exp\{-S_{0}(A_{\mu})
+\int dx[-iB(\partial^{\mu}A_{\mu})
+\bar{c}(-\partial_{\mu}\partial^{\mu})c]\}
\end{equation}
which agrees with (2.10) and (3.3).

We have thus shown that an identical physical meaning can be 
assigned to two Lagrangians (3.1) and (3.2) in suitably quantized
 theory.\\

In passing, a conventional way to relate (3.2) and (3.3) is to 
note
\begin{eqnarray}
&&\int {\cal D}A^{\omega}_{\mu}{\cal D}B{\cal D}\bar{c}{\cal D}c 
e^{\int dx [-\frac{1}{4}(\partial_{\mu}A_{\nu}-
\partial_{\nu}A_{\mu})^{2}-iB\partial_{\mu}A^{\omega}_{\mu}+ 
\bar{c}(-\partial_{\mu}\partial_{\mu})c]dx}\nonumber\\
&&=\int{\cal D}A^{\omega}_{\mu}{\cal D}B{\cal D}\bar{c}{\cal D}c 
e^{\int dx [-\frac{1}{4}(\partial_{\mu}A_{\nu}-
\partial_{\nu}A_{\mu})^{2}-\frac{1}{2}\xi B^{2}-
iB\partial_{\mu}A^{\omega}_{\mu}+ 
\bar{c}(-\partial_{\mu}\partial_{\mu})c]dx}\nonumber\\
&&=\int{\cal D}A^{\omega}_{\mu}{\cal D}\bar{c}{\cal D}c 
e^{\int dx [-\frac{1}{4}(\partial_{\mu}A_{\nu}-
\partial_{\nu}A_{\mu})^{2}-\frac{1}{2\xi}(\partial_{\mu}
A^{\omega}_{\mu})^{2}+ 
\bar{c}(-\partial_{\mu}\partial_{\mu})c]dx}
\end{eqnarray}
and by setting the gauge parameter $\xi=1$. The first equality
of (3.8), namely, the $\xi$ independence of the 
partition function is established by the BRST identity
as follows:
When one defines
\begin{equation}
Z(\xi)=\int{\cal D}A^{\omega}_{\mu}{\cal D}B{\cal D}\bar{c}
{\cal D}c 
e^{-S_{0}(A^{\omega})-\frac{\xi}{2}\int B^{2}dx +
\int[-iB
\partial_{\mu}
A^{\omega}_{\mu} +\bar{c}(-\partial_{\mu}\partial_{\mu})c]dx}
\end{equation}
with $S_{0}(A^{\omega})$ defined in (2.1),
one can show that 
\begin{eqnarray}
&&Z(\xi-\delta\xi)\nonumber\\
&&=\int{\cal D}A^{\omega}_{\mu}{\cal D}B{\cal D}\bar{c}{\cal D}c 
e^{-S_{0}(A^{\omega}_{\mu})-\frac{\xi-\delta\xi}{2}
\int B^{2}dx 
+\int[-iB
\partial_{\mu}
A^{\omega}_{\mu}+\bar{c}(-\partial_{\mu}\partial_{\mu})c]dx}
\nonumber\\
&&=Z(\xi)\\
&+&\int{\cal D}A^{\omega}_{\mu}{\cal D}B{\cal D}\bar{c}{\cal D}c 
(\frac{\delta\xi}{2}\int 
B^{2}dx)e^{-S_{0}(A^{\omega}_{\mu})-\frac{\xi}{2}\int B^{2}dx 
+\int[-iB
\partial_{\mu}
A^{\omega}_{\mu}+\bar{c}(-\partial_{\mu}\partial_{\mu})c]dx}.
\nonumber
\end{eqnarray}
On the other hand,the BRST invariance of the path integral 
measure and the effective action in the exponential factor gives
rise to  
\begin{eqnarray}
&&\int{\cal D}A^{\omega}_{\mu}{\cal D}B{\cal D}\bar{c}{\cal D}c 
[\int(\bar{c}B)dx]e^{-S_{0}(A^{\omega}_{\mu})-\frac{\xi}{2}
\int B^{2}dx 
+\int[-iB
\partial_{\mu}
A^{\omega}_{\mu}+\bar{c}(-\partial_{\mu}\partial_{\mu})c]dx}
\nonumber\\
&&=\int{\cal D}A^{\prime\omega}_{\mu}{\cal D}B^{\prime}{\cal 
D}\bar{c}^{\prime}{\cal D}c^{\prime} 
[\int(\bar{c}^{\prime}B^{\prime})dx]\nonumber\\
&&\times e^{-S_{0}(A^{\prime\omega}_{\mu})-\frac{\xi}{2}\int 
{B^{\prime}}^{2}dx 
+\int[-iB^{\prime}
\partial_
{\mu}A^{\prime\omega}_{\mu}+\bar{c}^{\prime}(-\partial_{\mu}
\partial
_{\mu})c^{\prime}]dx}\nonumber\\
&&=\int{\cal D}A^{\omega}_{\mu}{\cal D}B{\cal D}\bar{c}{\cal D}c 
[\int(\bar{c}B + \delta_{BRST}(\bar{c}B))dx]\nonumber\\
&&\times e^{-S_{0}(A^{\omega}_{\mu})-\frac{\xi}{2}\int B^{2}dx 
+\int[-iB\partial_{\mu}
A^{\omega}_{\mu}+\bar{c}(-\partial_{\mu}\partial_{\mu})c]dx}
\end{eqnarray}
where the BRST transformed variables are defined by 
$A^{\prime\omega}_{\mu}=A^{\omega}_{\mu}+
i\lambda\partial_{\mu}c,\\ B^{\prime} = B, 
\bar{c}^{\prime}=\bar{c}
+\lambda B, c^{\prime}=c $. The first equality of the above 
relation means that the path integral itself is independent of 
the naming of integration variables, and the second equality 
follows from the BRST invariance of the path integral measure
and the action. 
Namely, the BRST exact quantity vanishes as 
\begin{eqnarray}
&&\int{\cal D}A^{\omega}_{\mu}{\cal D}B{\cal D}\bar{c}{\cal D}c 
(\int\delta_{BRST}(\bar{c}B)dx)e^{-S_{0}(A^{\omega}_{\mu})
-\frac{\xi}{2}\int B^{2}dx 
+\int[-iB\partial_{\mu}A^{\omega}_{\mu}
+\bar{c}(-\partial_{\mu}\partial_{\mu})c]dx}
\nonumber\\
&&=
\int{\cal D}A^{\omega}_{\mu}{\cal D}B{\cal D}\bar{c}{\cal D}c 
(\lambda\int B^{2}dx)e^{-S_{0}(A^{\omega})-\frac{\xi}{2}
\int B^{2}dx 
+\int[-iB\partial_{\mu}A^{\omega}_{\mu}
+\bar{c}(-\partial_{\mu}\partial_{\mu})c]dx}
\nonumber\\
&&=0.
\end{eqnarray}
Thus from (3.10) we have the relation
\begin{equation}
Z(\xi-\delta\xi)=Z(\xi).
\end{equation}
Namely, $Z(\xi)$ is independent of $\xi$,
and $Z(1)=Z(0)$. This justifies the equality used in (3.8), and
we have established the conventional interpretation of (3.2) as 
an effective Lagrangian with a gauge fixing term added.
\\

Similarly, the two classical Lagrangians related to  
Yang-Mills fields
\begin{equation}
{\cal L}= -\frac{1}{4}(\partial_{\mu}A_{\nu}^{a}-
\partial_{\nu}A_{\mu}^{a}+gf^{abc}A_{\mu}^{b}A_{\nu}^{c})^{2}
-\frac{m^{2}}{2}A_{\mu}^{a}A^{a\mu}
\end{equation}
and 
\begin{equation}
{\cal L}= -\frac{1}{4}(\partial_{\mu}A_{\nu}^{a}-
\partial_{\nu}A_{\mu}^{a}+gf^{abc}A_{\mu}^{b}A_{\nu}^{c})^{2}
-\frac{1}{2}(\partial_{\mu}A^{a\mu})^{2}
\end{equation}
could be assigned an identical physical meaning as an effective 
gauge fixed Lagrangian associated with the quantum theory
defined by[1]
\begin{equation}
\int{\cal D}A_{\mu}^{a}{\cal D}B^{a}{\cal D}\bar{c}^{a}{\cal D}
c^{a}
\exp\{-S_{YM}(A_{\mu}^{a})
+\int dx[-iB^{a}(\partial^{\mu}A_{\mu}^{a})
+\bar{c}^{a}(-\partial_{\mu}(D^{\mu}c)^{a}]\}
\end{equation}
which is invariant under the quantum  gauge symmetry ( 
BRST transformation) with a Grassmann parameter $\lambda$
\begin{eqnarray}
&&\delta A^{a}_{\mu}=i\lambda(D_{\mu}c)^{a}\nonumber\\
&&\delta c^{a}=-\frac{i\lambda}{2}f^{abc}c^{b}c^{c}\nonumber\\
&&\delta\bar{c}^{a}=\lambda B^{a}\nonumber\\
&&\delta B^{a}=0.
\end{eqnarray}
In this analysis, we ignore the (ill-understood) Gribov-type
complications. This connection with the possible Gribov-type
complications becomes transparent if one considers 
\begin{equation}
f(A^{a\mu})=\frac{1}{2}(\partial_{\mu}A^{a\mu})^{2}
\end{equation}
in (3.15). We then obtain in the modified scheme (1.1) and its
equivalent formula (1.3) (by suppressing 
the Yang-Mills indices)
\begin{equation}
D^{\mu}\frac{\delta f(A_{\nu})}
{\delta A_{\mu}}=
D_{\mu}\partial^{\mu}(\partial_{\nu}A^{\nu})
\end{equation}
and the associated determinant factor contains the operator
\begin{equation}
D_{\mu}\partial^{\mu}(\partial_{\nu}D^{\nu})
+\frac{\delta D_{\mu}(A_{\mu}^{\omega})}{\delta\omega}|_{\omega=0}
\partial^{\mu}(\partial_{\nu}A^{\nu}).
\end{equation}
The gauge fixing and Faddeev-Popov terms then become in the 
modified scheme
\begin{eqnarray}
{\cal L}_{g}&=&-iBD_{\mu}\partial^{\mu}(\partial_{\nu}A^{\nu})
-\bar{c}[D_{\mu}\partial^{\mu}(\partial_{\nu}D^{\nu})
+\frac{\delta D_{\mu}(A_{\mu}^{\omega})}{\delta\omega}|_{\omega=0}
\partial^{\mu}(\partial_{\nu}A^{\nu})]c\nonumber\\
&=&-iBD_{\mu}\partial^{\mu}(\partial_{\nu}A^{\nu})
-\bar{c}D_{\mu}\partial^{\mu}(\partial_{\nu}D^{\nu})c.
\end{eqnarray}
In this last step, we used the fact that the gauge fixing 
condition 
\begin{equation}
D_{\mu}\partial^{\mu}(\partial_{\nu}A^{\nu})=0
\end{equation}
is equivalent to 
\begin{equation}
\partial_{\nu}A^{\nu}=0
\end{equation}
in Euclidean theory if the Gribov-type complications are
absent and thus the inverse of the operator 
$D_{\mu}\partial^{\mu}$ is well-defined. We thus have the 
path integral formula in the modified scheme
\begin{eqnarray}
&&\int{\cal D}A_{\mu}^{a}{\cal D}B^{a}{\cal D}\bar{c}^{a}{\cal D}
c^{a}\nonumber\\
&&\ \ \ \ \times\exp\{-S_{YM}(A_{\mu}^{a})
+\int dx[-iBD_{\mu}\partial^{\mu}(\partial^{\nu}A_{\nu})
-\bar{c}D_{\mu}\partial^{\mu}(\partial_{\nu}D^{\nu})c\}
\nonumber\\
&&=\int{\cal D}A_{\mu}^{a}{\cal D}B^{a}{\cal D}\bar{c}^{a}{\cal D}
c^{a}\nonumber\\
&&\ \ \ \ \times\exp\{-S_{YM}(A_{\mu}^{a})
+\int dx[-iB(\partial^{\mu}A_{\mu})
+\bar{c}(-\partial_{\mu}(D^{\mu}c)]\}
\end{eqnarray}
after the re-definition of auxiliary variables
\begin{equation}
BD_{\mu}\partial^{\mu} \rightarrow B, \ \ \ 
\bar{c}D_{\mu}\partial^{\mu}\rightarrow \bar{c}
\end{equation}
which leaves the path integral measure invariant. 
This last re-definition is allowed only when the 
operator $D_{\mu}\partial^{\mu}$ is well-defined, namely, in the 
absence of Gribov-type complications in Euclidean theory.\\

We have illustrated that the apparent ``massive gauge
field'' in the classical level (3.14) has no intrinsic 
physical meaning. It can be interpreted either as a  
massive (non-gauge) vector theory, which is a conventional 
interpretation, or as a gauge-fixed effective 
Lagrangian for a massless gauge field, which is allowed in the 
modified quantization scheme.
In the framework of path integral quantization, these different 
{\em interpretations} of a given classical Lagrangian could also 
be understood that  we have a certain freedom 
in the choice of the path integral measure (although the 
conventional interpretation of the measure is perfectly possible
as is shown in (1.5)): One choice of the 
measure 
\begin{eqnarray}
&&\int d\mu\exp\{\int dx[-\frac{1}{4}(\partial_{\mu}A_{\nu}^{a}-
\partial_{\nu}A_{\mu}^{a}+gf^{abc}A_{\mu}^{b}A_{\nu}^{c})^{2}
-\frac{m^{2}}{2}A_{\mu}^{a}A^{a\mu}]\}\nonumber\\
&&\equiv\int{\cal D}A_{\mu}\frac{1}{\int{\cal 
D}g\exp[-\int\frac{m^{2}}{2}(A^{a g}_{\mu})^{2}dx]}\\
&&\times\exp\{\int dx[-\frac{1}{4}(\partial_{\mu}A_{\nu}^{a}-
\partial_{\nu}A_{\mu}^{a}+gf^{abc}A_{\mu}^{b}A_{\nu}^{c})^{2}
-\frac{m^{2}}{2}A_{\mu}^{a}A^{a\mu}]\}\nonumber
\end{eqnarray}
gives rise to a renormalizable gauge theory (3.16), and the other 
choice,which is suggested by the naive  classical analysis,
\begin{eqnarray}
&&\int d\mu\exp\{\int dx[-\frac{1}{4}(\partial_{\mu}A_{\nu}^{a}-
\partial_{\nu}A_{\mu}^{a}+gf^{abc}A_{\mu}^{b}A_{\nu}^{c})^{2}
-\frac{m^{2}}{2}A_{\mu}^{a}A^{a\mu}]\}\\
&&\equiv\int{\cal D}A_{\mu}
\exp\{\int dx[-\frac{1}{4}(\partial_{\mu}A_{\nu}^{a}-
\partial_{\nu}A_{\mu}^{a}+gf^{abc}A_{\mu}^{b}A_{\nu}^{c})^{2}
-\frac{m^{2}}{2}A_{\mu}^{a}A^{a\mu}]\}\nonumber
\end{eqnarray}
gives rise to a massive {\em non-gauge} vector theory, which is
not renormalizable. 
As we emphasized in Section 1, we here allow a more general
definition of path integral measure  than the one suggested by 
the naive classical analysis on the basis of Dirac brackets, if
one should take (3.14) as a fundamental Lagrangian. 
It is 
interesting that the choice of the measure in (3.26) 
defines a renormalizable unitary theory starting with the 
apparently inconsistent theory (3.14). It is an advantage
of the modified quantization scheme\cite{zwanziger}
\cite{jona-lasinio} that we can readily assign a physical 
meaning to a wider class of theories.
A somewhat analogous situation to (3.26) arises when one attempts
 to quantize the so-called anomalous gauge theory: A suitable 
choice of the measure with a Wess-Zumino term gives rise to a 
consistent quantum theory in 2-dimensions, for example
\cite{jackiw}. 

From a view point of classical-quantum correspondence, one can 
define a classical theory uniquely starting from quantum theory 
by considering the limit $\hbar\rightarrow 0$, but not the other 
way around in general. For example, the ambiguities related to 
the operator ordering are well known in any quantization, though 
the present choices of path integral measure are not directly 
related to operator ordering ambiguities.

In the context of the present broader interpretation of classical
 massive gauge fields, the massive gauge fields generated by the 
Higgs mechanism are exceptional and quite different.
The Higgs mechanism for Abelian theory, for example, is defined
by (in this part, we use the Minkowski metric with $g_{\mu\nu}
=(1,-1,-1,-1)$)
\begin{equation}
{\cal L}=(D^{\mu}\phi)^{\dagger}D_{\mu}\phi -\mu^{2}|\phi|^{2}-
\lambda|\phi|^{4}-\frac{1}{4}(\partial_{\mu}A_{\nu}-
\partial_{\nu}A_{\mu})^{2}
\end{equation}
which is manifestly gauge invariant with $D_{\mu}=\partial_{\mu}
-igA_{\mu}$. The mass $m=gv$ for the 
gauge field is generated after the spontaneous symmetry 
breaking of gauge symmetry defined by 
\begin{equation}
\phi(x)\equiv\phi^{\prime}(x)+v/\sqrt{2}
\end{equation}
with
\begin{equation}
v^{2}=-\mu^{2}/\lambda
\end{equation}
for $\mu^{2}<0$.
In this procedure, all the terms in the Lagrangian including 
the mass term generated by the Higgs mechanism are gauge 
invariant. Consequently, our argument discussed so far (i.e,.
a possible re-interpretation of the mass term as a gauge 
fixing term ) does not apply to the present massive vector 
particle whose mass is generated by the Higgs mechanism. It is 
quite satisfactory that the Higgs mechanism has an intrinsic 
physical meaning even in our extended interpretation of mass 
terms.  

In view of the well known fact that the massive non-Abelian gauge
theory is inconsistent as a quantum theory (3.27), it may be 
sensible to treat all the classical massive non-Abelian 
Lagrangians as a gauge fixed version of pure non-Abelian gauge
 theory and to restrict the massive non-Abelian gauge fields to 
those generated by the Higgs mechanism. In this connection,
we note that our discussion is quite different from the 
Stueckelberg formalism of the classical massive vector theory,
which was also an attempt to make sense out of a massive
vector theory by introducing extra scalar freedom. In the 
Stueckelberg formalism, one attempts to understand the 
classical massive vector theory as a quantum massive vector 
theory, whereas our consideration here proposes to assign a 
physical meaning to a classical massive vector theory as a gauge 
fixed version of a massless gauge theory.

\section{Dynamical generation of gauge fields}

It is a long standing question if one can generate gauge 
fields from some {\em more} fundamental mechanism. In fact,
there have been numerous attempts in the past to this effect.
To our knowledge, however, there exists no definite convincing
scheme so far. On the contrary, there is a no-go theorem or 
several arguments against such an 
attempt\cite{case}\cite{weinberg}. We here briefly comment on 
this issue from a view point of  our extended scheme of quantum
gauge symmetry; our comment in this section is 
inevitably quite speculative.

Apart from technical details, the basic argument against the 
``dynamical'' generation of gauge fields is that the  Lorentz
invariant positive definite theory cannot simply generate the 
negative metric states associated with the time components of 
massless gauge fields. In contrast, the dynamical generation 
of the Lagrangian of the structure 
\begin{equation}
{\cal L}=-\frac{1}{4}(\partial_{\mu}A_{\nu}^{a}-
\partial_{\nu}A_{\mu}^{a}+gf^{abc}A_{\mu}^{b}A_{\nu}^{c})^{2}
-f(A_{\mu}^{a})
\end{equation}
does not appear to be prohibited by general arguments so far.
Here the term $f(A_{\mu}^{a})$ is Lorentz invariant but not 
invariant under the local gauge symmetry and thus breaks the 
gauge symmetry explicitly; 
$f(A_{\mu}^{a})=\frac{m^{2}}{2}(A_{\mu}^{a})^{2}$ is the simplest 
and most attractive example, since it carries the lowest scaling
dimension. The appearance of $f(A_{\mu}^{a})$, which is not
gauge invariant, is also natural if one starts with a fundamental
 theory without gauge symmetry.

Here comes the issue of interpretation of the induced 
Lagrangian (4.1). If one regards (4.1) as a quantum theory 
from the beginning, what one generates is simply a non-gauge 
theory: This is also the case if one evaluates a general 
S-matrix, which effectively represents the Lagrangian (4.1), and 
looks for the possible poles corresponding to massless gauge 
particles. 

However, one might consider that 
the induced Lagrangian such as (4.1) is a {\em classical} 
object which should be quantized anew: In terms of path integral 
language, the Lagrangian is induced when one integrates over
the ``fundamental'' degrees of freedom, and one need to perform 
further path integral over the induced Lagrangian anew.
If one takes this latter view point, one might be allowed to
 regard the part of $f(A_{\mu}^{a})$, which breaks classical 
gauge symmetry, as a gauge fixing term in the modified 
quantization scheme\cite{zwanziger}\cite{jona-lasinio}. In
 this latter interpretation, one 
might be allowed to say that massless gauge fields are generated 
dynamically. Although a dynamical generation of pure gauge 
fields is prohibited, a {\em gauge fixed} Lagrangian might be 
allowed to be generated. (In this respect, one may recall that 
much of the arguments for the no-go 
theorem\cite{case}\cite{weinberg} would be  refuted if one could 
generate a gauge fixed Lagrangian with the Faddeev-Popov term 
added.) 
 The mass for the gauge field which has 
an intrinsic unambiguous physical meaning is then further 
induced by the spontaneous symmetry breaking of the gauge 
symmetry thus defined (the Higgs mechanism). 
 
We next comment on a mechanism for generating gauge fields by 
the violent random fluctuation of gauge degrees of freedom at 
the beginning of the universe\cite{nielsen}; this scheme
is based on the renormalization group flow starting from an 
initial chaotic theory. In such a scheme, it is natural to think 
that one is always dealing with quantum theory, and thus no room 
for our way of re-interpretation of the induced theory.
Nevertheless, we find a possible connection in the following
sense: To be precise, an example of massive Abelian gauge 
field in {\em compact} lattice gauge theory
\begin{equation}
\int{\cal D}U\frac{{\cal D}\Omega}{vol(\Omega)}
\exp[-S_{inv}(U)-S_{mass}(U^{\Omega})]
\end{equation}
is analyzed in Ref.\cite{nielsen}. 
Here $S_{inv}(U)$ stands for the gauge invariant part of 
the lattice Abelian gauge field $U$, and $S_{mass}(U^{\Omega})$
stands for the gauge non-invariant mass term with the gauge 
freedom $\Omega$. In compact theory, one need not fix the 
gauge and instead one may take an average over the entire gauge 
volume of $\Omega$.
They argued that the mass term, which breaks gauge symmetry
softly, disappears in the long distance limit when one integrates 
over the entire gauge freedom $\Omega$. Their scheme is apparently
dynamical one, in contrast to the kinematical nature of our 
re-interpretation. Nevertheless, the massive Abelian theory is a 
free theory in continuum formulation, and the disappearance of 
the mass term by a mere smearing over the gauge volume may suggest
 that the mass term in their scheme is also treated as a kind of
gauge artifact, just as in our kinematical re-interpretation. 

In passing, we note that  the lattice gauge theory by using a 
mass term as a gauge fixing term has been discussed in the 
context of a numerical simulation\cite{golterman}. The main 
interest there is to evaluate gauge dependent quantities such as 
the propagator. 

\section{Discussion}

On the basis of the equivalence between (1.1) and the local 
expression (1.3), we 
have discussed a possible more flexible interpretation of 
classical Lagrangians.  In this broader interpretation of 
the classical action, the quantum gauge symmetry (BRST symmetry) 
is  defined for a much wider class of theories than pure 
classical gauge theory, such as Maxwell field and Yang-Mills 
fields, in suitable quantization. In this framework, 
classical gauge symmetry is sufficient to generate 
quantum gauge symmetry (up to quantum anomalies), but it is not 
necessary in general; a theory, whose gauge symmetry is broken 
by some terms in the Lagrangian, could be re-interpreted as being
gauge fixed by those terms  in 
suitably  quantized theory.

In this broader interpretation, the BRST symmetry is quite 
universal. This universality 
presumably arises from the fact that the essence of BRST symmetry
 is quite simple; geometrically, it is defined as 
the translation and  scale transformations of a superspace 
coordinate specified by the real element of the Grassmann 
algebra\cite{fujikawa2}
\begin{eqnarray}
&&Q: \theta\rightarrow \theta 
+ \lambda,\ \ \ (BRST \ \ charge)\nonumber\\
&&D: \theta\rightarrow e^{\alpha}\theta,\ \ \ 
(ghost\ \ number\ \ charge)
\end{eqnarray}
where $\theta$ and $\lambda$ are the real elements of the 
Grassmann algebra and $\alpha$ is a real number. Namely, the 
abstract BRST symmetry by itself carries no information of 
classical gauge symmetry.

We hope that the observation in the present note 
will stimulate further thinking on the meaning of classical and 
quantized gauge fields and also on the possible
origin of gauge fields, the most profound notion of modern 
field theory.\\

The present paper is a revised and extended  version of 
our contribution to ICHEP2000 in Osaka in 
July, 2000.
One of us (KF) thanks H. Sugawara for a comment on a classical
limit of quantized theory.

\end{document}